\documentclass[runningheads,a4paper]{llncs}

\usepackage{aliascnt}
\usepackage{tabu}
\usepackage{amssymb,amsmath,mathtools}
\usepackage{booktabs}
\usepackage{multirow}
\usepackage{natbib}

\usepackage{times}
\usepackage{url}
\usepackage{color}
\usepackage{multirow}
\usepackage{multicol}
\usepackage{graphicx}
\usepackage{caption}
\usepackage{tikz}
\usetikzlibrary{calc,tikzmark}
\usepackage[shortcuts]{extdash}
\renewcommand{\cite}[1]{\citep{#1}}
\usepackage[ruled,norelsize,linesnumbered]{algorithm2e}

\title{Learning Transformations of Musical Material using Gated Autoencoders}
\author{Stefan Lattner, Maarten Grachten, Gerhard Widmer}

\institute{Department of Computational Perception\\
  Johannes Kepler University\\
  Linz, Austria}

\authorrunning{Stefan Lattner, Maarten Grachten, Gerhard Widmer}

\begin{document}

\maketitle

\begin{abstract}
Music is usually highly structured and it is still an open question how to design models which can successfully learn to recognize and represent musical structure.
A fundamental problem is that structurally related patterns can have very distinct appearances, because the structural relationships are often based on \emph{transformations} of musical material, like chromatic or diatonic transposition, inversion, retrograde, or rhythm change.
In this preliminary work, we study the potential of two unsupervised learning techniques---Restricted Boltzmann Machines (RBMs) and Gated Autoencoders (GAEs)---to capture pre-defined transformations from constructed data pairs.
We evaluate the models by using the learned representations as inputs in a discriminative task where for a given type of transformation (e.g. diatonic transposition), the specific relation between two musical patterns must be recognized (e.g. an upward transposition of diatonic steps).
Furthermore, we measure the reconstruction error of models when reconstructing musical transformed patterns.
Lastly, we test the models in an analogy-making task.
We find that it is difficult to learn musical transformations with the RBM and that the GAE is much more adequate for this task, since it is able to learn representations of specific transformations that are largely content-invariant.
We believe these results show that models such as GAEs may provide the basis for more encompassing music analysis systems, by endowing them with a better understanding of the structures underlying music.
\end{abstract}

\section{Introduction}\label{sec:introduction}
An important notion in western music is that of structure: the phenomenon that a musical piece is not an indiscriminate stream of events, but can be decomposed into temporal segments, often in a hierarchical manner.
An important factor determining what we regard as structural units is the relation of these units to each other.
The most obvious relation is the literal repetition of a segment of music on multiple occasions in the piece.
However, more complex music tends to convey structure not only through repetition, but also through other types of relations between structural units, such as chromatic or diatonic transpositions, or rhythmical changes.
We refer to these relations in general as \emph{transformations}.

\begin{figure}[t]
\includegraphics[trim=1cm 21.5cm 1cm 1cm, clip, width=1.\linewidth]{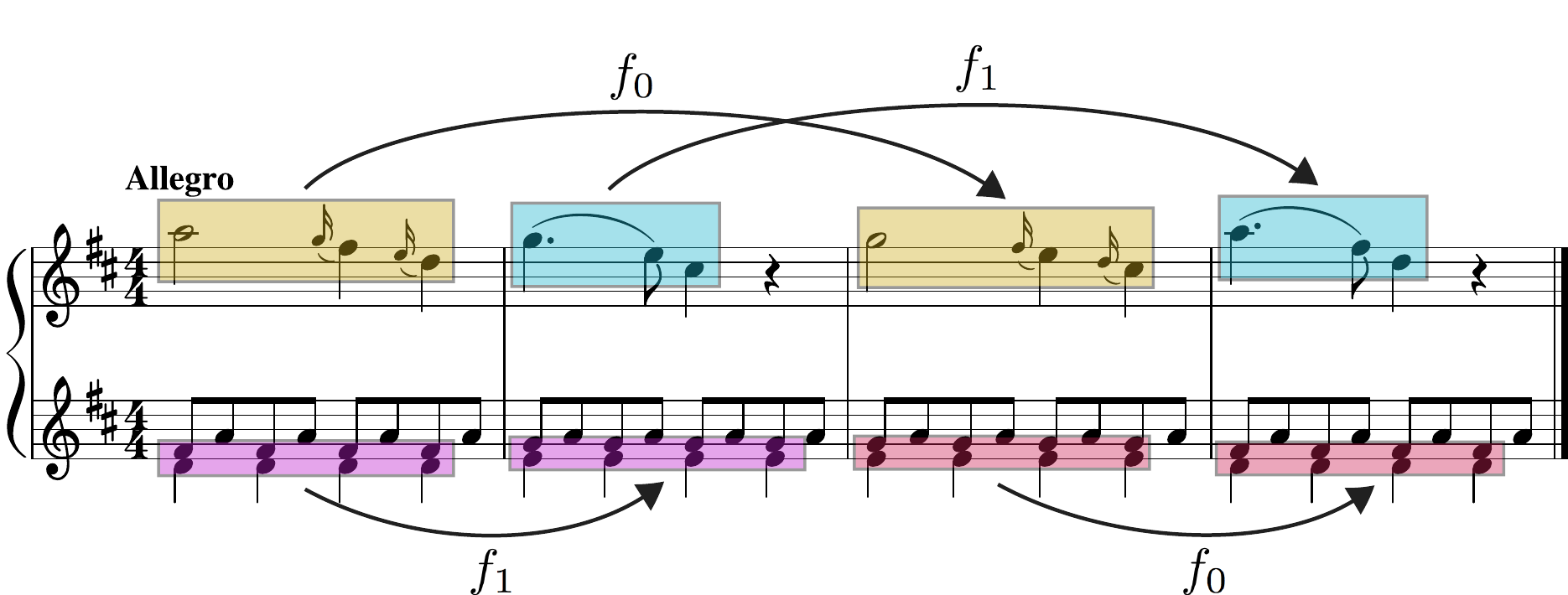} \\
\vspace{-3mm} \\
\includegraphics[trim=.1cm 0cm .6cm 0cm, clip, width=1.\linewidth]{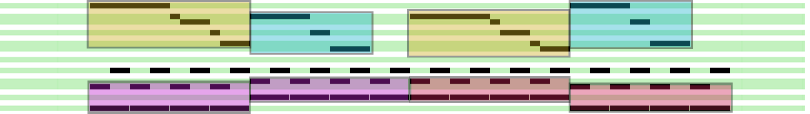}
\caption{Beginning of the ``Rondo in D major'' by W.
A.
Mozart in Western music notation and in a piano roll representation, where transformation functions $f_n$ constitute diatonic transpositions (best viewed in color).
Green horizontal lines mark the diatonic pitches in the scale of D major.}
\label{fig:rondo}
\end{figure}

Figure~\ref{fig:rondo} shows an example of mid-level structure induced by transformations in an excerpt of a rondo by W.~A.~Mozart where related phrases are marked by boxes in the same color.
The yellow boxes mark the input and output of a function $f_0$ performing a diatonic transposition by -1 scale step.
The same function is applied in the bass section, marked with red boxes.
Likewise, $f_1$ performs a diatonic transposition by +1 scale step in the melody section (blue boxes) and in the bass section (purple boxes).
The transformations defined in $f_0$ and $f_1$ constitute structural relationships which may be applied to any musical material and in that sense are \emph{content-invariant}.
Note how this view on music as a collection of basic musical material that is transformed in a variety of ways can provide very concise and schematically simple representations of a musical piece.
Finding such a representation may be a goal in itself (music analysis), or it may serve as the basis for other tasks, such as music summarization, music classification and similarity estimation, or computer-assisted composition.

There are some fundamental challenges to automated detection of structure in music however.
Firstly, although there are a few common types of transformations such as chromatic and diatonic transposition, it is not easy to give an exhaustive list of transformation types that should be considered when comparing potential structural units of a musical piece.
Ideally, rather than relying on human intuition, we would like to infer transformations from actual music that can account for a large portion of structure in that music.
A second challenge is that the definition of musical structure in terms of transformations between structural units is descriptive, not prescriptive.
Composers may use transformation as a compositional tool, but in their artistic freedom they may perform further ad-hoc alterations at will. 
The resulting musical material may thus exhibit \emph{approximate} transformations, deviating from exact transformations in subtle but unpredictable ways.

State of the art methods for motivic analysis of music, such as the compression-based methods discussed in~\cite{louboutin16:_using}, do not directly address these issues.
We believe that a more data-driven approach is needed, where transformations are learned from data in an unsupervised manner.
This naturally leads us to consider connectionist models for unsupervised learning, such as restricted Boltzmann machines (RBMs)~\cite{hinton06}, and autoencoders~\cite{bengio-2009}.
A particularly promising approach towards unsupervised learning of content-invariant musical transformations are bilinear models like Gated Autoencoders (GAEs) \cite{memisevic2011gradient} or Factored Boltzmann Machines \cite{memisevic2010learning}.
By employing \emph{multiplicative interactions}, these models can learn typical correlations between pairs of data instances.
It was shown that GAEs are effective at encoding transformations between image pairs, such as spatial translation and rotation of objects \cite{memisevic2013learning}.

The purpose of this paper is not to show how such models can be used to perform music analysis, but (rather more modestly) to investigate their ability to learn transformations from given pairs of musical fragments, more specifically n-grams of unit-sized vertical slices from a pianoroll notation.
To this end, we define common musical transformations (chromatic transposition, diatonic transposition, tempo change, and retrograde) and construct n-gram pairs accordingly (see Section~\ref{sec:Experiment}). Such a controlled setting provides ground-truth data and allows for performing a proof-of-concept on the general suitability of GAEs in music, and should be regarded as a first step towards identifying and learning transformations implicit in a music corpus.


We compare a GAE with a standard RBM in learning the aforementioned pre-defined musical transformations between n\=/gram pairs of symbolic polyphonic music. We test the discriminative performance of the models in a classification task (see Section~\ref{sec:discuss_discriminative}) and their expressivity based on the reconstruction cross-entropy (see Section~\ref{sec:discuss_recon}).
Furthermore, we test the performance of the GAE in applying learned relations to data instances not seen during training (see Section~\ref{sec:discuss_gen}).

In Section~\ref{sec:related-work} we discuss related models and concepts, and in Section~\ref{sec:method} we describe the GAE and the RBM models used in our experiments.
The experiment setup including data preparation, used model architectures, and training is introduced in Section~\ref{sec:Experiment}.
Results are presented and discussed in Section~\ref{sec:results-discussion} and Section~\ref{sec:concl-future-work} gives an outlook on future work and challenges.

\section{Related work}\label{sec:related-work}
The problem of detecting musical relations falls in the class of \emph{analogy-making}  \cite{hofstadter95}, an important capability of the human brain in which the objective is to produce a data instance X given the three instances A, B, C, and the query “X is to A as B is to C”.
\citet{nichols2012musicat} shows how analogy-making also plays an important role in music cognition, and ``Musicat'', a musical analogy-making model is presented.
Identifying analogies between musical sections exhibiting the same transformation is a first step towards identifying similarities between transformed musical objects (i.e.
by utilizing \emph{transformation-invariant} representations \cite{memisevic2013aperture}).

The problem of relating data instances is also tackled by some deep-learning methods, like in \cite{reed2015deepanalogy}, where visual analogy-making is performed by a deep convolutional neural network using a supervised strategy.
Siamese architectures are used by \citet{bromley1993signature} for signature verification, and by \citet{chopra2005learning} to identify identical persons from face images in different poses.
With a GAE, an explicit mapping between two inputs is learned.
In contrast, with common deep-learning methods relations are often indirectly qualified by comparing representations of related instances.
For example, it is a common approach to construct a space in which operations like addition and subtraction of data vectors imply some semantic meaning \cite{mikolov2013distributed,pennington2014glove}.

An analogy-making model similar to the GAE is the Transforming Autoencoder introduced by \citet{Hinton:2011wm}.
In contrast to the GAE this model is supervised, as during training the respective transformations have to be specified in a parameterized way (e.g.
rotation angle or distance of shift in image transformation).
Another related model is the Spatial Transformer Network, which transforms an input image (conditioned on itself) before passing it on to classification layers \cite{jaderberg2015spatial}.

Bilinear models themselves are used for analogy-making on rotating 3D objects \cite{memisevic2013aperture}, for modeling facial expression \cite{susskind2011modeling}, to learn transformation-invariant representations for classification tasks \cite{ICML2012Memisevic_105}, and for time-series prediction on accelerated movements and sine-sweeps by stacking more than one GAE and learning higher-order derivatives \cite{michalski2014modeling}.

The GAE is based on a method introduced by \citet{adelson1985spatiotemporal}, where motion patterns in the three-dimensional $x$-$y$-$t$ space are modeled by filter pairs receptive to distinct orientations in that space.
It shows that mapping units in a GAE function similar to \emph{complex cells} in the visual cortex, which gives rise to important perceptual processes like detecting fluent motion from series of static images.

\section{Method}\label{sec:method}

\subsection{Gated Autoencoder} \label{sec:autoencoder}
\begin{figure}[t]
\begin{center}
\includegraphics[width=.5\linewidth]{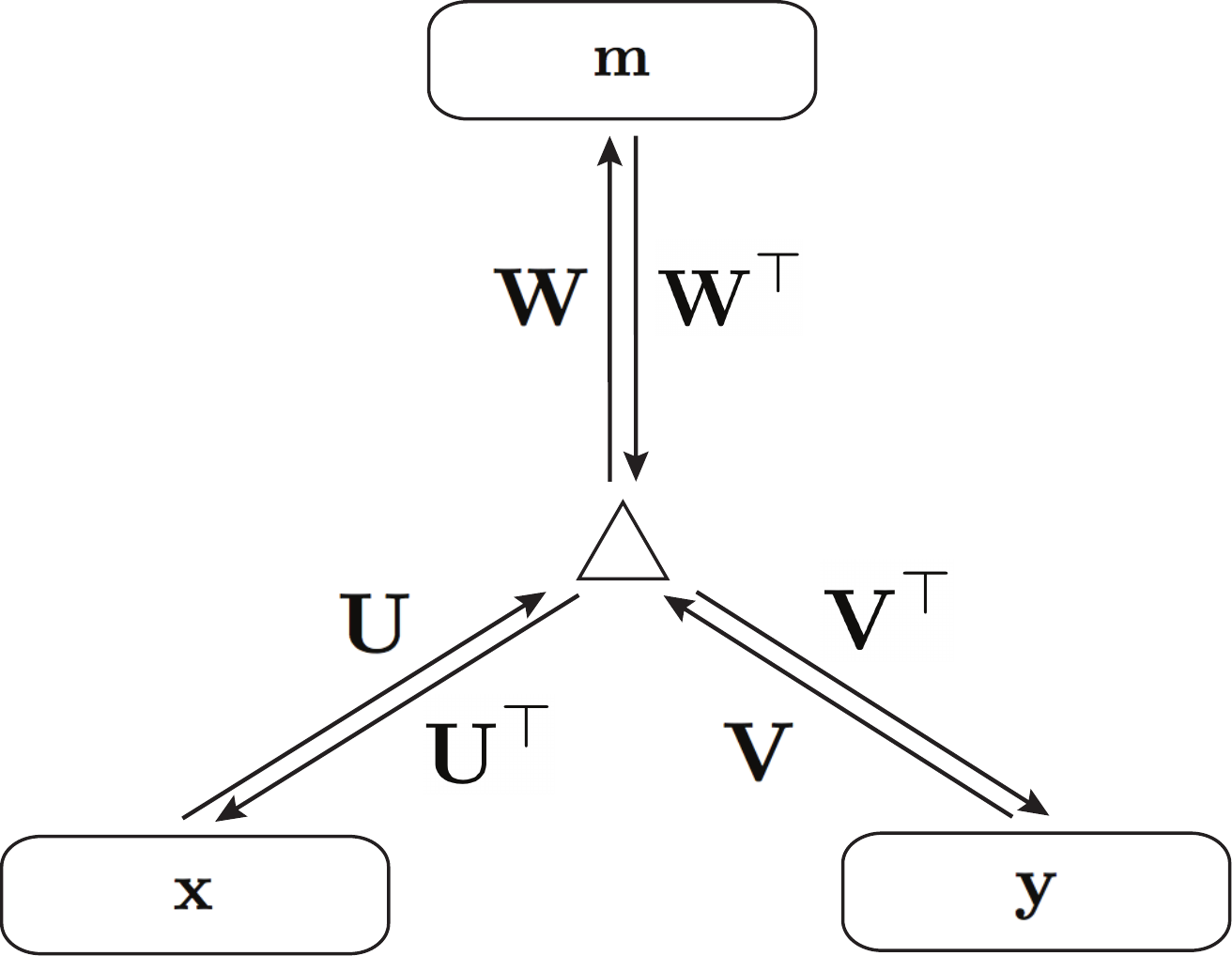}
\caption{Schematic illustration of a Gated Autoencoder.}
\label{ga}
\end{center}
\end{figure}

\begin{figure}
\begin{center}
\includegraphics[width=\linewidth]{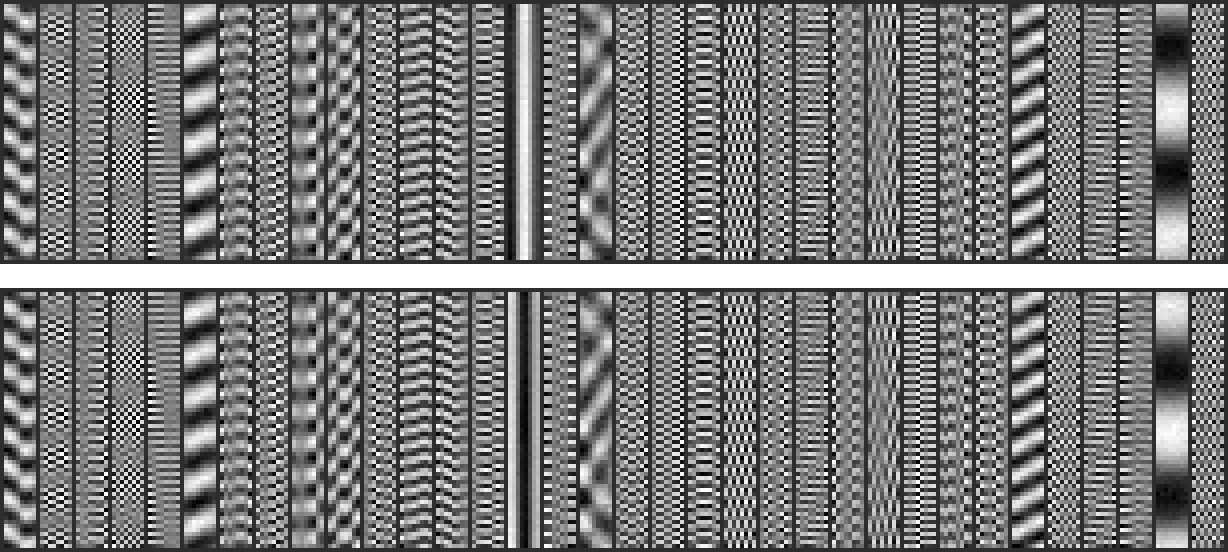}
\end{center}
\caption{Some complementary filters of $\mathbf{U}$ (top) and $\mathbf{V}$ (bottom) learned from \emph{transposed} pairs of musical n\=/grams.}
\label{fig:filters}
\end{figure}

A Gated Autoencoder (GAE) learns mappings between data pairs $\mathbf{x},\mathbf{y}\in\mathbb{R}^{P}$ using lateral mapping units $\mathbf{m}\in\mathbb{R}^{L}$ as
\begin{equation}\label{eq:gamap}
\mathbf{m} = \sigma (\mathbf{W}(\mathbf{Ux} \cdot \mathbf{Vy})),
\end{equation}
where $\mathbf{U}, \mathbf{V} \in \mathbb{R}^{P \times O}$ and $\mathbf{W} \in \mathbb{R}^{O \times L}$ are weight matrices, and $\cdot$ denotes the element-wise product (see Figure~\ref{ga} for an illustration).
In our experiments, $\sigma$ is the sigmoid function, but other non-linearities, like the square-root or softplus are also reported in the literature \cite{adelson1985spatiotemporal,alain2013gated,ICML2012Memisevic_105}.
The element-wise (i.e.
pairwise) product between filter responses $\mathbf{Ux}$ and $\mathbf{Vy}$, leads to filter pairs in $\mathbf{U}$ and $\mathbf{V}$ encoding \emph{correspondences} between the inputs $\mathbf{x}$ and $\mathbf{y}$.
GAEs are very expressive in representing transformations, with the potential to represent any permutation in the input space (''shuffling pixels'') \cite{memisevic2013learning}.
The resulting filter pairs show transformation-specific features, like phase-shifted Fourier components when learning transposition between n\=/grams of polyphonic music in our experiments (see Figure~\ref{fig:filters}).
Such phase-shifted filter pairs are then receptive to corresponding shifts between pairs of data instances.

An interesting aspect of a GAE is its symmetry when solving Equation \ref{eq:gamap} for its inputs, allowing for a reconstruction of an input given the other input and a mapping code.
The reconstruction for binary data is given by

\begin{equation}\label{recon1}
\mathbf{\tilde{x}} = \sigma (\mathbf{U^\top} (\mathbf{W}^\top \mathbf{m} \cdot \mathbf{Vy})),
\end{equation}
and likewise
\begin{equation}\label{recon2}
\mathbf{\tilde{y}} = \sigma (\mathbf{V}^\top (\mathbf{W}^\top \mathbf{m} \cdot \mathbf{Ux})).
\end{equation}
A commonly used training loss \cite{memisevic2013learning,michalski2014modeling} is the symmetric reconstruction error
\begin{equation}\label{eq:recon_symm}
\mathcal{L} = ||\mathbf{x} - \mathbf{\tilde{x}}||^2 + ||\mathbf{y} - \mathbf{\tilde{y}}||^2.
\end{equation}

\begin{figure}
\begin{tabular}{cc}
\includegraphics[trim=31mm 25mm 25mm 25mm, clip, width=.45\linewidth]{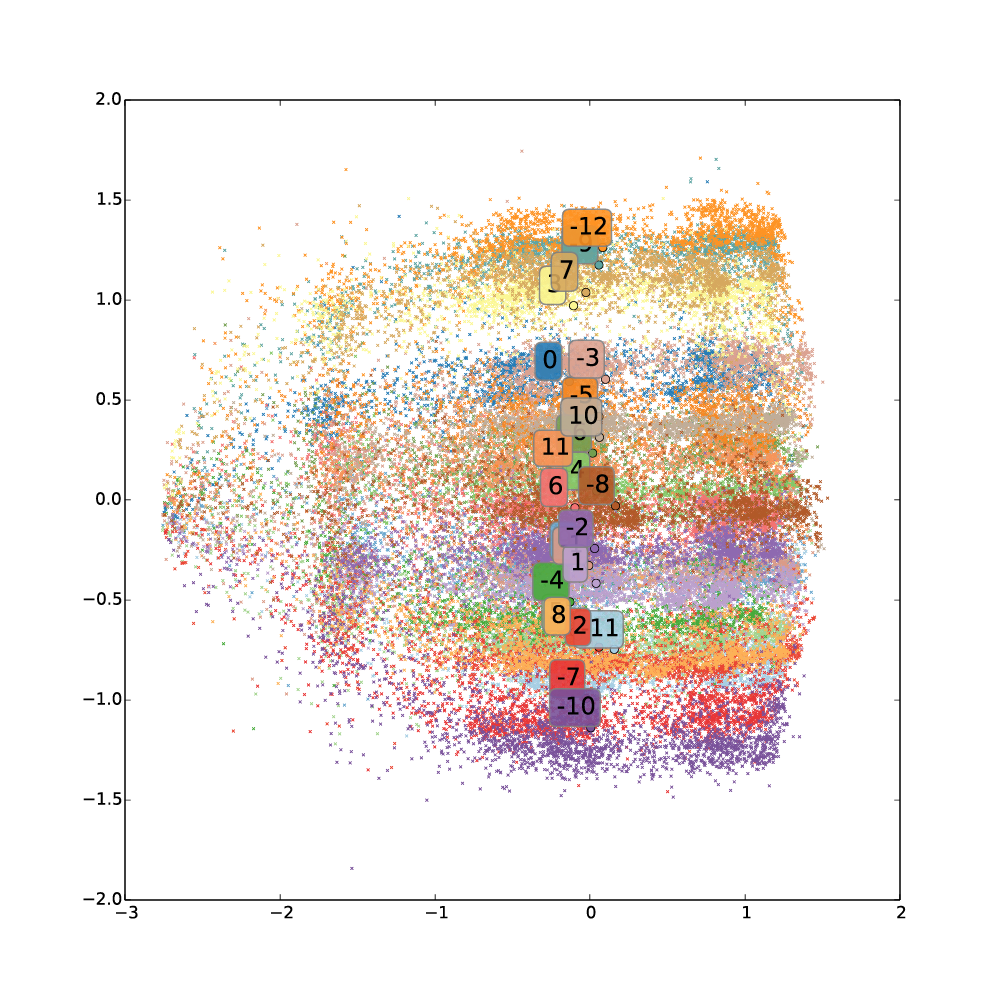} &
\includegraphics[trim=31mm 25mm 25mm 25mm, clip, width=.45\linewidth]{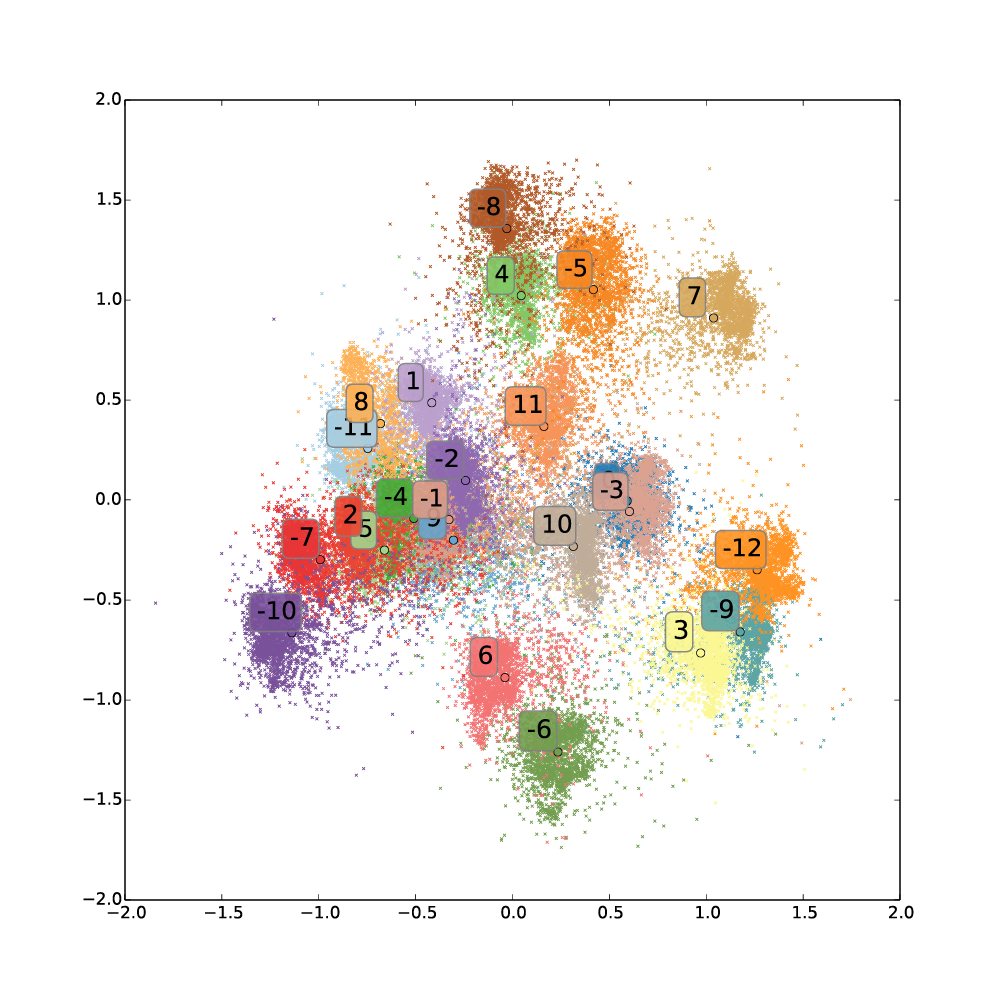} \\
a & b \\
\end{tabular}
\caption{First two principal components (a) and second and third principal component (b) of mappings resulting from unsupervised learning of pairs exhibiting the \emph{chromatic transposition} (TransC) relation.
Points are colored and cluster centers are named according to the different classes within the TransC relation type (i.e.
distances of note shifts).}
\label{pca}
\vspace{-0.2cm}
\end{figure}

When using additive interactions in neural networks, units resemble logical OR-gates which accumulate evidence.
In contrast, multiplicative interactions resemble AND-gates that can detect co-occurrences.
Multiplicative interactions enable the model to ignore input that does not exemplify a known transformation, making it less content-dependent \cite{memisevic2013learning}.
Ideally, learned representations encode fully content-invariant transformations between data pairs.
In practice, there is always some content-dependence in the resulting codes, and it is a future challenge to tackle this issue (see Figure~\ref{pca}, where the variance along the first principal component is content-dependent, as it is not correlated with transformation classes).

\subsection{Restricted Boltzmann Machine} \label{sec:rbm}
An RBM is an energy-based stochastic neural network with two layers, a visible layer with units $\mathbf{v} \in \{0,1\}^r$ and a hidden layer with units $\mathbf{h} \in \{0,1\}^q$ \cite{Hinton:2002ic}.
The units of both layers are fully interconnected with weights $\mathbf{W} \in \mathbb{R}^{r \times q}$, while there are no connections between the units within a layer.
Given a visible vector $\mathbf{v}$, the free energy of the model is calculated as

\begin{equation}\label{equ:energy}
\mathcal{F}(\mathbf{v}) = - \mathbf{a}^\intercal\mathbf{v} - \sum_i \log \left( 1 + e^{\left(b_i + \mathbf{W}_i \mathbf{v} \right)}\right),
\end{equation}

\noindent where $\mathbf{a} \in \mathbb{R}^r$ and $\mathbf{b} \in \mathbb{R}^q$ are bias vectors, and $\mathbf{W}_i$ is the $i$-th row of the weight matrix.
The model is trained by minimizing the free energy given the training data, and by maximizing the free energy for samples from the model---a method called Contrastive Divergence \cite{Hinton:2002ic}.

Reconstruction is performed by projecting a data instance into the hidden space, followed by a projection of the resulting hidden unit activations back into the input space. The reconstruction cross-entropies--as presented in Section \ref{sec:discuss_recon}--arise from the error between the input before and after the projection.

\section{Experiment}\label{sec:Experiment}
The aim of the current experiment is to test the performance of a GAE in learning musical relations between pairs of n\=/grams.
As a baseline, we train an RBM with approximately the same number of parameters on \emph{concatenated} pairs of n\=/grams, like in \cite{susskind2011modeling}.
In order to have a controlled setting and labels for the transformation classes, we construct pairs where one item is an original n\=/gram selected at random from a set of polyphonic Mozart sonatas (see Figure \ref{results_generate}(A) for n\=/gram examples), and the other item is constructed by applying a randomly selected transformation to this n\=/gram.
The experiment is unsupervised in that there are no targets or labels when training the GAE, but we only present one type of transformation per training session.
The goal of the GAE is to cluster the classes \emph{within} each type of transformation.
For example, within the transformation type \emph{chromatic transposition} (TransC), the classes are the different semitone intervals by which the items in a pair are shifted with respect to each other.

We use the resulting representations as an input to a classification feed-forward Neural Network (FFNN), in order to assess their discriminativity with respect to the given classes.
Furthermore, we measure the reconstruction cross-entropies for both models, and we apply transformations extracted from single examples to new data.

\subsection{Data}\label{sec:TrainingData}
We use MIDI files encoding the scores of Mozart piano sonatas from the Mozart/Batik data set~\cite{widmer2003discovering}.
From those pieces, we create binary piano-roll representations, using the MIDI pitch range $[36,100]$ and a temporal resolution of 1/16th note.
From those representations we create $8$-grams (i.e. any consecutive eight 1/16th slices from the piano-roll representation) and for each $8$-gram we construct a counterpart which exhibits the respective transformation.
In order to remove potential peculiarities of the corpus with respect to absolute shifts and keys, we start by shifting the notes of the original n\=/grams by $k$ semitones taken randomly from the set $[-12,11]$.
In the following, we describe how the respective counterparts are constructed.

\subsubsection{Chromatic Transposition (TransC)}
Chromatic transposition is performed by shifting every pitch in a set of notes by a fixed number of semitones (cf.
Figure \ref{results_generate}(1)).
For each n\=/gram, we construct a counterpart by shifting all notes by $k$ semitones taken from the set $[-12,11]$.
Note that this also includes the transposition by $0$ semitones, which is the special case ``exact repetition''.
The resulting $24$ classes for testing the discriminative performance of our models are consequently the elements of the set $[-12,11]$.

\subsubsection{Diatonic Transposition (TransD)}
Diatonic transposition is the transposition within a key.
This is, any pitch in a set of notes is shifted by a fixed number of \emph{scale steps} (i.e.
transposition using only allowed notes, e.g.
depicted as green lines in Figure \ref{fig:rondo}).
This operation may change the intervals between notes (see Figure~\ref{results_generate}(2) for examples of diatonic transposition).
At first, we estimate the key for each n\=/gram by choosing the one with the least accidentals from the set of keys for which all pitches of the n\=/gram are ``allowed'', omitting n\=/grams which do not fit in a key.
Using this method, we obtain n\=/grams which are assigned to a unique key and can be unambiguously transformed into another key.
That way, a unique transformation can be learned for any constructed instance pair.
We create counterparts for each n\=/gram by shifting each pitch by $k$ \emph{scale steps} of the estimated scale, taken from the set $[-7,6]$.
Thus, the resulting $14$ classes the models are trained on are the elements of the set $[-7,6]$.

\subsubsection{Half time / double time (Tempo)}
''Half time'' means doubling the durations of all notes, and ``double time'' is the inverse operation.
We realize the half time relation by scaling each n\=/gram to double width and then taking only the first half of the result (cf.
Figure \ref{results_generate}(3)), and for the inverse operation we swap such pairs.
Consequently, we assess the performance of the models in discriminating between the relations \{double time, half time, identity\}.

\subsubsection{Retrograde (Retro)}
In music, retrograde means to reverse the temporal order of given notes.
We can simply create pairs exhibiting the retrograde relation by flipping n\=/grams horizontally.
This results in two classes: \{retrograde, identity ($\neg$retrograde)\}


\subsection{Model Architectures}
\subsubsection{Architecture of the GAE and the RBM}
In order to obtain comparable results for the GAE and the RBM, for each test we choose the same number of parameters and layers for each model.
The GAE may be seen as a two-layered model, where the factors constitute the hidden units of the first layer, and the mapping units constitute the hidden units of the second layer.
Consequently, we also use a two-layered RBM architecture where the number of hidden units in the first layer is equal to the number of factors in the GAE, and the number of hidden units in the top layer is equal to the number of mapping units in the GAE.
We test three different model sizes for every type of transformation.
The smallest model (128/64) has 128 units in the first layer and 64 units in the second layer.
The next bigger model has 256/128 units, and the biggest model has 512/256 units in the respective layers.

\subsubsection{Architecture of the Classification FFNN}
For all classification tasks, the same architecture is used: A three-layered feed-forward Neural Network (FFNN) with $512$ Rectified Linear units (ReLUs) in the first layer, $256$ ReLUs in the second layer, and Softmax units in the output layer.
The size of the output layer is equal to the number of classes.

\begin{table*}
\centering
\scriptsize
\setlength{\tabcolsep}{.5em}
\renewcommand{\arraystretch}{1.3}
\begin{tabular}{rrrrrrrrrrrrrrrrr}
\toprule
& & \multicolumn{14}{c}{predicted interval} & $\sum$\\
& & -7 & -6 & -5 & -4 & -3 & -2 & -1 & 0 & 1 & 2 & 3 & 4 & 5 & 6 &$\text{diag}$ \\
\midrule
\multirow{15}{-2em}{\rotatebox{90}{target interval}} & -7 & 7187 & 1 & \tikzmark{e1}1 & 1 & 3 & \tikzmark{c1}3 & 0 &  \tikzmark{a}10 & 1 & \tikzmark{c2}2 & 1 & 0 & 4 & 0 &  \\
& -6 & 6 & 6943 & 2 & 14 & 2 & 7 & 9 & 1 & 30 & 1 & 30 & 0 & 7 & 6 & \textbf{0} \\
& -5 & \tikzmark{e2}15 & 21 & 7043 & 14 & 26 & 13 & 6 & 3 & 3 & 29 & 0 & 6 & 10 & 3 & \textbf{10} \\
& -4 & 1 & 10 & 3 & 7055 & 7 & 5 & 0 & 0 & 6 & 1 & 17 & 0 & 5 & 0 & \textbf{10} \\
& -3 & 2 & 1 & 6 & 55 & 7090 & 6 & 11 & 0 & 2 & 3 & 8 & 34 & 1 & 9 & \textbf{11} \\
& -2 & \tikzmark{c4}5 & 4 & 3 & 16 & 11 & 6960 & 10 & 2 & 4 & 0 & 12 & 2 & 20 & 0 & \tikzmark{d2}\textbf{52} \\
& -1 & 1 & 8 & 1 & 1 & 6 & 3 & 7050 & 5 & 6 & 3 & 6 & 7 & 13 & 24 & \textbf{3} \\
& 0 & \tikzmark{a1}7 & 0 & 2 & 0 & 0 & 4 & 0 & 7124 & 0 & 6 & 0 & 3 & 3 & 0 & \tikzmark{b}\textbf{164} \\
& 1 & 0 & 46 & 3 & 14 & 7 & 5 & 18 & 6 & 7028 & 0 & 23 & 1 & 4 & 13 & \textbf{28} \\
& 2 & \tikzmark{c3}13 & 1 & 25 & 3 & 16 & 8 & 3 & 0 & 2 & 7082 & 15 & 16 & 4 & 2 & \tikzmark{d1}\textbf{59} \\
& 3 & 1 & 5 & 0 & 12 & 10 & 9 & 5 & 2 & 5 & 0 & 6972 & 39 & 6 & 1 & \textbf{33} \\
& 4 & 1 & 2 & 6 & 0 & 12 & 3 & 6 & 2 & 3 & 7 & 19 & 7027 & 1 & 1 & \textbf{29} \\
& 5 & 9 & 7 & 5 & 17 & 0 & 89 & 11 & 8 & 7 & 4 & 48 & 11 & 6874 & 35 & \tikzmark{f1}\textbf{117} \\
& 6 & 2 & 21 & 5 & 4 & 17 & 6 & 41 & 1 & 36 & 6 & 7 & 23 & 4 & 7053 & \textbf{135} \\
\multicolumn{3}{l}{\rule{0pt}{2.5ex}$\sum\text{diag}$} &\textbf{2} & \textbf{30} & \textbf{13} & \textbf{12} & \tikzmark{d3}\textbf{58} & \textbf{7} & \tikzmark{b1}\textbf{232} & \textbf{32} & \tikzmark{d4}\textbf{104} & \textbf{42} & \textbf{30} & \tikzmark{f2}\textbf{158} & \textbf{141} & \textbf{98488} \\
\bottomrule
\end{tabular}
\begin{tikzpicture}[overlay, remember picture, red, yshift=.25\baselineskip, shorten >=.5pt, shorten <=.5pt]
\draw [->] ($(pic cs:a)+(0pt,3pt)$) -- ($(pic cs:b)+(0pt,5pt)$);
\draw [->] ($(pic cs:a1)+(0pt,4pt)$) -- ($(pic cs:b1)+(0pt,8pt)$);
\end{tikzpicture}
\begin{tikzpicture}[overlay, remember picture, blue, yshift=.25\baselineskip, shorten >=.5pt, shorten <=.5pt]
\draw [->] ($(pic cs:c1)+(0pt,3pt)$) -- ($(pic cs:d1)+(0pt,3pt)$);
\draw [->] ($(pic cs:c4)+(0pt,3pt)$) -- ($(pic cs:d4)+(0pt,8pt)$);
\end{tikzpicture}
\begin{tikzpicture}[overlay, remember picture, orange, yshift=.25\baselineskip, shorten >=.5pt, shorten <=.5pt]
\draw [->] ($(pic cs:c2)+(0pt,3pt)$) -- ($(pic cs:d2)+(0pt,4pt)$);
\draw [->] ($(pic cs:c3)+(0pt,5pt)$) -- ($(pic cs:d3)+(0pt,6pt)$);
\end{tikzpicture}
\begin{tikzpicture}[overlay, remember picture, green, yshift=.25\baselineskip, shorten >=.5pt, shorten <=.5pt]
\draw [->] ($(pic cs:e1)+(0pt,3pt)$) -- ($(pic cs:f1)+(0pt,4pt)$);
\draw [->] ($(pic cs:e2)+(0pt,5pt)$) -- ($(pic cs:f2)+(0pt,8pt)$);
\end{tikzpicture}
\vspace{0.3cm}
\footnotesize
\caption{Confusion Matrix for classifier FFNN trained on representations of GAE TransD 128/64.}
\vskip-2ex
\label{results_confusion}
\vspace{-0.3cm}
\end{table*}

\subsection{Training}
In the following, we report how the GAE, the RBM, and the FFNN are trained in our experiments.
For each transformation type, we train on $490\, 000$ pairs using a validation set of $10\, 000$ pairs.
The final testing is done on a test set of size $100\, 000$ per transformation type.

\subsubsection{Training the Gated Autoencoder}
The model parameters of the GAE are trained using stochastic gradient descent to minimize the symmetric reconstruction error (see Equation \ref{eq:recon_symm}).
We train the model on the data pairs for $1000$ epochs, using a mini-batch size of $500$, a learning rate of $3\times10^{-5}$ and a momentum of $0.93$.
Learning improves when the input (i.e. $\mathbf{y}$ and $\mathbf{x}$ in Equations \ref{recon1} and \ref{recon2}, respectively) is corrupted during training, as it is done in de-noising autoencoders \cite{vincent2010stacked}. We achieve this by randomly setting $35\%$ of the input bits to zero and train the GAE to reconstruct an uncorrupted, transformed version of it (i.e. $\mathbf{\tilde{x}}$ and $\mathbf{\tilde{y}}$ in Equations \ref{recon1} and \ref{recon2}, respectively). For the first $100$ epochs, the input weights are re-scaled after each parameter update to their average norm, as described by \citet{susskind2011modeling,memisevic2011gradient,ICML2012Memisevic_105}. We use L1 and L2 weight regularization on all weights, and Lee's sparsity regularization \cite{Lee:2007uz} on the mapping units and on the factors.

\subsubsection{Training the Restricted Boltzmann Machine}
We train two RBM layers on \emph{concatenated} data pairs with greedy layer-wise training using \emph{persistent contrastive divergence} (PCD)~\cite{tielemanPcd2008}, a variation of the standard \emph{contrastive divergence} (CD) algorithm~\cite{Hinton:2002ic}, which is known to result in a better approximation of the likelihood gradient.
We use a learning rate of $3 \times 10^{-3}$ which we reduce to zero during $300$ training epochs.
We use a batch size of $100$, and we reset (i.e.
randomize) the weights of neurons, whose average activity over all training examples exceeds $85\%$. We use L1 and L2 weight regularization and the sparsity regularization proposed by~\citet{Goh:2010wi} setting $\mu = 0.08$ and $\phi = 0.75$.

\subsubsection{Training the Classification FFNN}
The FFNN is trained in a supervised manner on the representations (i.e. the neural activation patterns) resulting from unsupervised training of the GAE and the RBM on related data pairs, using the categorical cross-entropy loss.
The network is trained for $300$ epochs with a learning rate of $0.005$, a batch size of $100$ and a momentum of $0.93$.
We apply L2 weight regularization on all weights, sparsity regularization as in \cite{Lee:2007uz}, $50\%$ dropout \cite{srivastava2014dropout} on all except the top-most layer, as well as batch normalization \cite{ioffe2015batch}.

\section{Results and discussion}\label{sec:results-discussion}


\subsection{Discriminative performance}\label{sec:discuss_discriminative}
Table~\ref{results_class} shows the mis-classification rates of the FFNN, trained on representations of the GAE and the RBM.
The models are trained separately for each transformation type, chromatic transposition (TransC), diatonic transposition (TransD), half-time/double-time (Tempo), and retrograde (Retro) where training data pairs of each type contain several type-specific classes (see Section~\ref{sec:TrainingData}).

\begin{table}[t]
\centering
\footnotesize
\begin{tabu}{rllll}
\toprule
& \multicolumn{4}{c}{Transformation} \\ \cline{2-5}
\rule{0pt}{2.5ex}  Model  \ \ \ \ &  TransC  \ \  \ \ & TransD  \ \ \ \ & Tempo  \ \ \ \ & Retro  \ \ \\ \midrule
\rowfont{\footnotesize}
 \multicolumn{1}{l}{\ \ \# Classes  \ \ \ \ \rule{0pt}{2ex}} & 24 & 14 & 3 & 2 \\
 \multicolumn{1}{l}{\ \ \textbf{Random \rule{0pt}{3ex}}} & 95.83 & 92.86 & 66.67 & 50.00 \\
 \multicolumn{1}{l}{\ \ \textbf{RBM \rule{0pt}{3ex}}\ \ \ \ \ \ } & & & & \\
 \ \ 128/64  \ \  \ \ & 23.10 & 87.55 & 50.09 & 50.06  \\
 \ \ 256/128  \ \  \ \ & 6.12 & 51.05 & 47.52 & 50.11  \\
 \ \ 512/256  \ \  \ \ & 2.18 & 19.47 & 40.33 & 50.19  \\
 \multicolumn{1}{l}{\ \ \textbf{GAE} \rule{0pt}{2.5ex}} & & & & \\
 \ \ 128/64  \ \  \ \ & 1.88 & 1.51 & 2.47 & 3.24 \\
 \ \ 256/128  \ \  \ \ & \textbf{0.02} & 0.26 & 0.89 & 1.10 \\
 \ \ 512/256  \ \  \ \ & 0.03 & \textbf{0.23} & \textbf{0.11} & \textbf{0.28} \\
\bottomrule
\end{tabu}
\vspace{0.3cm}
\caption{Mis-classified rates (in percent) from the classification Feed-Forward Neural Network trained on representations of the Restricted Boltzmann Machine (RBM) and the Gated Autoencoder (GAE) in different architecture sizes and for different transformation types.
\emph{Random} denotes the random guessing baseline dependent on the number of classes.}
\label{results_class}
\vspace{-0.4cm}
\end{table}

The GAE is clearly better at learning discriminative representations of musical relations.
The mis-classification rates of the largest architecture (512/256) are below $0.3\%$ for all relations.
In contrast, the RBM is less suitable for unsupervised learning of relations--in the case of Retro it does not even outperform random guessing.
Note furthermore that in contrast to the GAE, increasing the capacity of the RBM does not always improve results.
This suggests that the architecture itself is a limiting factor:
As the activations of units in an RBM only depend on additive accumulation of evidence, it attempts to learn all combinations of mutually transformed data instances it is presented with during training. For transposition relations, learning such combinations results in representations which are informative enough that the classification FFNN can infer the transformation class from it. Retrograde and Tempo, however, are too complex transformations to grasp for content-dependent representation learner.

In contrast, the GAE separates the problem in representing characteristics of the input n-grams with additive input connections on the one hand, and modeling transformations using multiplicative interactions on the other hand. In addition, when training the GAE on reconstructing an input, the other input provides content information allowing the mapping units to represent only the content-invariant transformation. It is still a non-trivial result that the GAE is able to learn all pre-defined transformation types. In particular diatonic transposition (TransD) is a very relevant transformation in music and it is an interesting finding that the GAE is capable of learning it.

Table~\ref{results_confusion} shows a confusion matrix for the GAE TransD 128/64 relation.
To interpret the table, it is instructive to realize that two diatonic transpositions of the same ngram yield another pair of n-grams that are also diatonically transposed versions of each other.
For instance, the diatonic transposition pairs 0 and -7,  1 and -6, 2 and -5, and so on, all yield a resultant transposition of -7 (a downward octave).
Analog equivalences hold for resultant transpositions of upward octave, the upward/downward fifth, third, and so on.
The equivalent resultant transpositions have been marked by colored arrows in the table.
The arrows clearly highlight that the model is biased toward confusions \emph{by one octave} from the correct shift (red arrows).
For example, when the actual shift between n\=/gram pairs is $+5$, the classifier frequently estimates a shift of $-2$, which is $7$ scale steps or \emph{one octave} away from the correct target.
Similarly, frequently confused targets are a third (green arrows), a fifth (blue arrows), and an octave plus a third (orange arrows) away from the correct target.
Since the data pairs are uniformly distributed over all classes, it follows that these confusions are caused by covariances \emph{within} single n\=/grams, where thirds, fifths, and octaves occur frequently. The mis-classifications of second interval shifts (i.e. the entries directly above and below to the diagonal) most likely result from second-order covariances. For example, shifts by -3 and -4 are sometimes confused because their sum is -7, which is again an octave.

\subsection{Reconstruction}\label{sec:discuss_recon}
Table~\ref{results_recon} lists the reconstruction cross-entropies for each architecture versus transformation type.
The error of the largest GAE architecture is about an order of magnitude smaller than that of the best RBM architecture.
Furthermore, the cross-entropies do not substantially improve with increasing capacity of the RBM, again suggesting an architectural advantage of the GAE over the RBM.
\citet{olshausen2007bilinear} suggest that the advantage consists in the ability to separate content and its manifestation (the ``what'' and the ``where'' ).
Mapping units only have to encode the "where" components leading to a more efficient use of the model parameters.

\begin{table}[t]
\vspace{-.27cm}
\centering
\footnotesize
\begin{tabular}{rlllll}
\toprule
& \multicolumn{4}{c}{Transformation} \\ \cline{2-5}
\rule{0pt}{2.5ex}  Model  \ \  \ \ & TransC \ \  \ \  & TransD \ \  \ \  & Tempo \ \  \ \  & Retro \ \ \\ \midrule
\multicolumn{1}{l}{ \ \  \textbf{RBM \rule{0pt}{2ex}}\ \ \ \ \ \ \ \ } & & & & \\
128/64  \ \  \ \ & 0.131 & 0.122 & 0.098 & 0.110 \\
256/128  \ \  \ \  & 0.122 & 0.119 & 0.084 & 0.095 \\
512/256 \ \  \ \  & 0.113 & 0.112 & 0.075 & 0.086 \\
\multicolumn{1}{l}{\ \ \textbf{GAE} \rule{0pt}{2.5ex}} & & & & \\
128/64 \ \  \ \  & 0.026 & 0.032 & 0.025 & 0.033 \\
256/128 \ \  \ \  & 0.018 & 0.024 & 0.015 & 0.017 \\
512/256 \ \  \ \  & \textbf{0.012} & \textbf{0.016} & \textbf{0.007} & \textbf{0.009} \\
\bottomrule
\end{tabular}
\vspace{0.3cm}
\caption{Reconstruction cross-entropies (per input unit) for the Restricted Boltzmann Machine (RBM) and the Gated Autoencoder (GAE) in different sizes and for different transformation types.}
\label{results_recon}
\vspace{-0.6cm}
\end{table}

\subsection{Generation}\label{sec:discuss_gen}
\begin{figure}[!ht]
\newlength{\figheight}
\setlength{\figheight}{.395\linewidth}
\setlength{\textfloatsep}{-0cm}
\begin{tabular}{cc|c}
\rule{0pt}{3ex} & A & B \\
1 & \raisebox{-.5\figheight}{\rule{0em}{18ex}\includegraphics[height=\figheight]{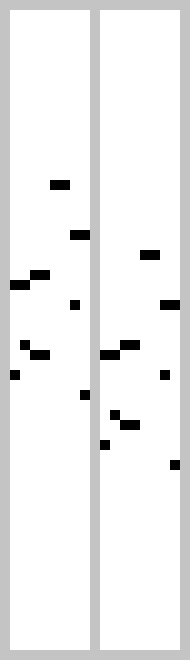}} & \raisebox{-.5\figheight}{\includegraphics[height=\figheight]{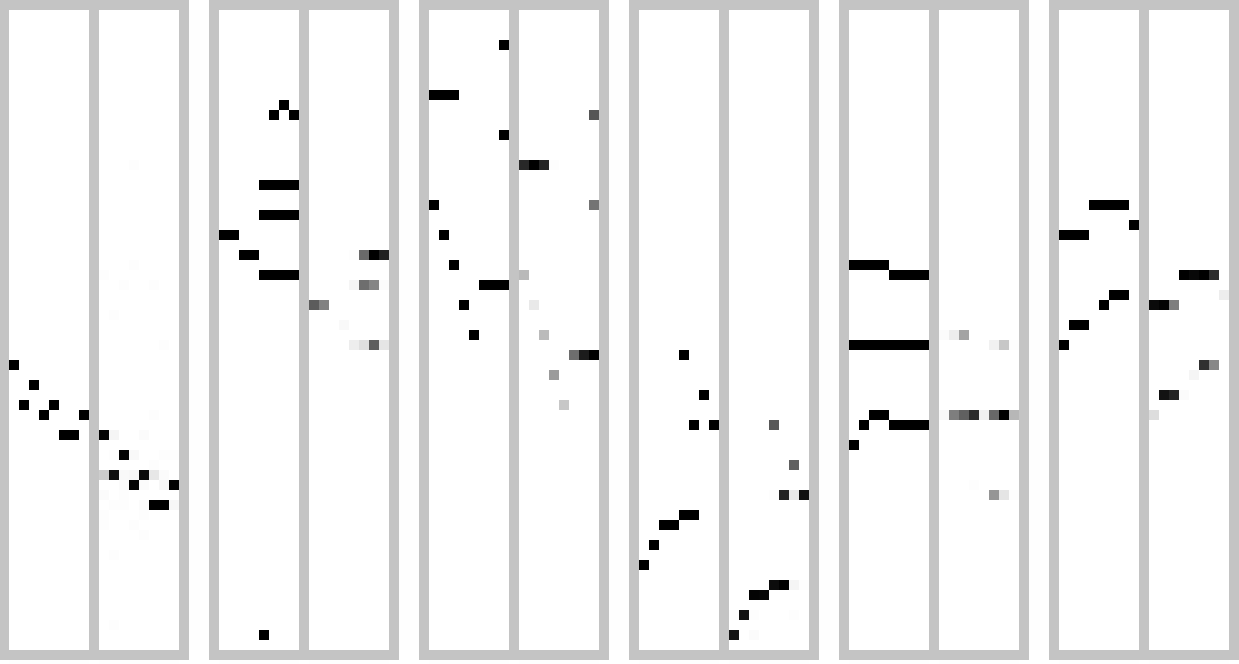}}\\
2 & \raisebox{-.5\figheight}{\rule{0em}{22ex}\includegraphics[height=\figheight]{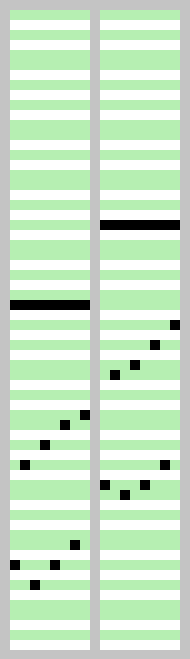}} & \raisebox{-.5\figheight}{\includegraphics[height=\figheight]{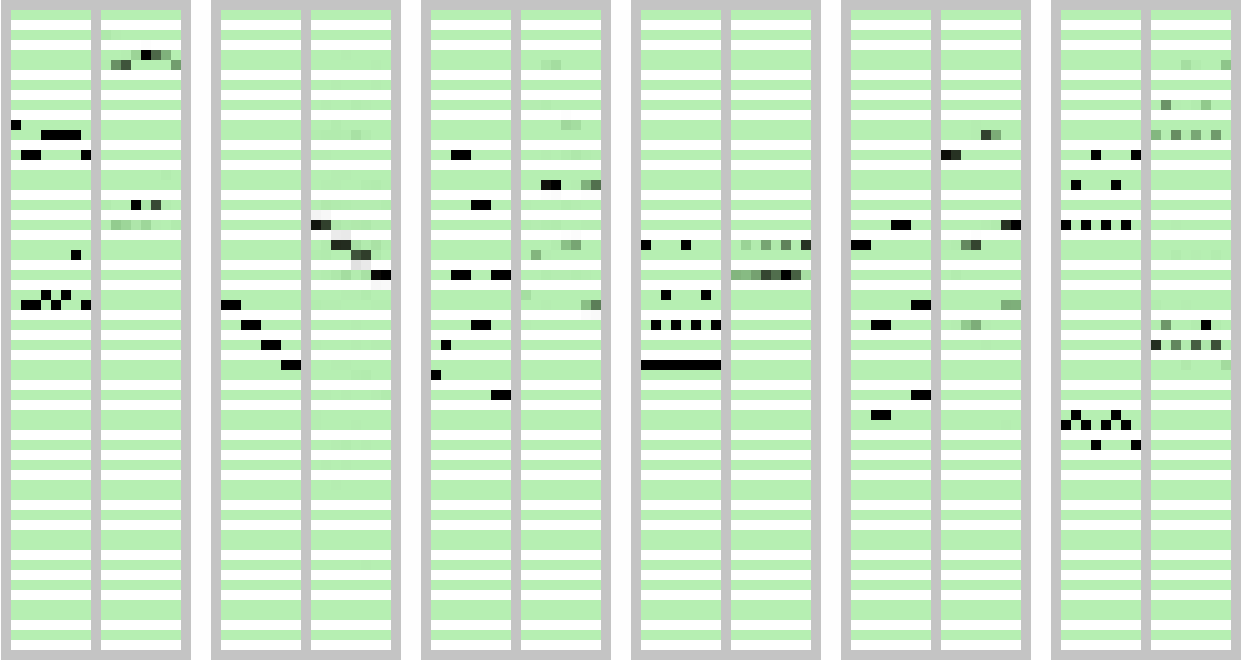}}\\
3 & \raisebox{-.5\figheight}{\rule{0em}{22ex}\includegraphics[height=\figheight]{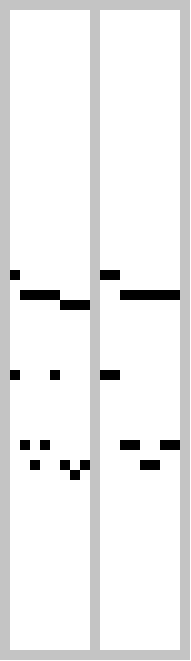}} & \raisebox{-.5\figheight}{\includegraphics[height=\figheight]{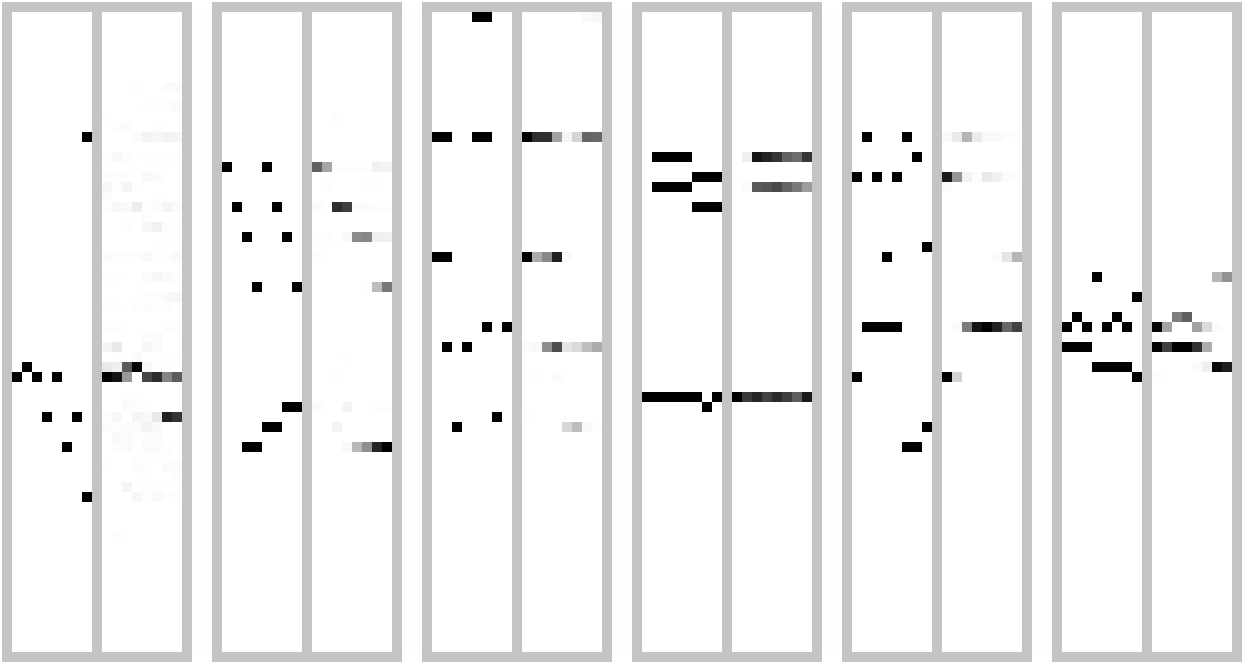}}\\
4 & \raisebox{-.5\figheight}{\rule{0em}{22ex}\includegraphics[height=\figheight]{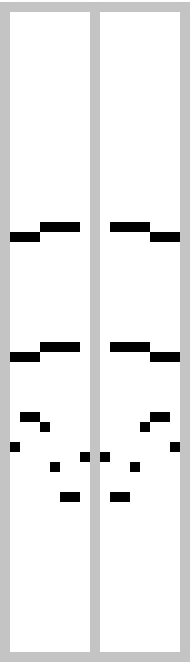}} & \raisebox{-.5\figheight}{\includegraphics[height=\figheight]{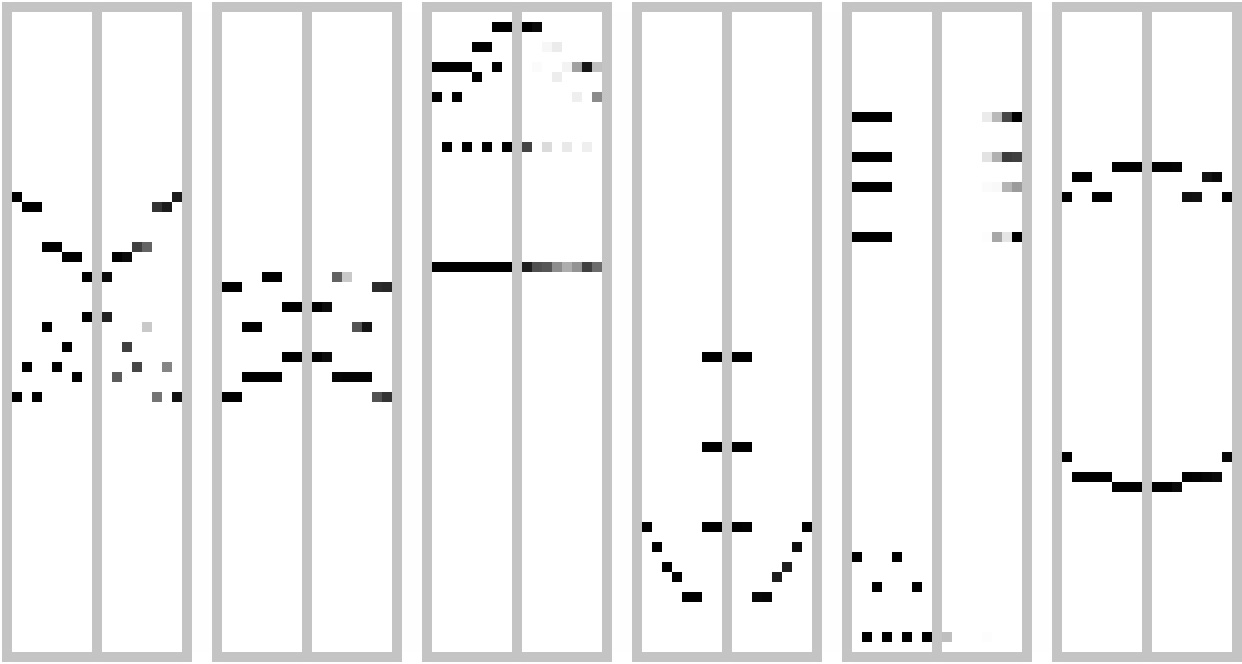}}\\
\end{tabular}
\caption{Results of the analogy-making task. Transformations are inferred by the GAE from pairs of 8-grams (A) and applied to new n-grams, not seen during training (B, left parts) to generate counterparts with analogous transformations (B, right parts), where the level of blackness indicates the certainty of the model that the respective note is part of the transformed result: 1) Chromatic transposition, 2) diatonic transposition, 3) tempo change (halftime), and 4) retrograde.
Green horizontal lines mark scale notes in diatonic transposition, for visual guidance.
}
\label{results_generate}
\vspace{-0.2cm}
\end{figure}

Figure~\ref{results_generate} shows analogy-making examples.
Mappings are inferred from template n\=/gram pairs (A) and applied to single instances which were not part of the training corpus.
The resulting pairs (B) should exhibit the same transformation as the template pairs from which the transformation was inferred.
For diatonic transposition (2), the mapping is applied only to instances in the same key as the source instance (i.e.
left instance) of the template pairs, and valid pitches are marked with green lines.

The transformations of the template pairs shown in Figure~\ref{results_generate} are chromatic transposition by $-7$ semitones (1), diatonic transposition by $+5$ scale steps in C major (2), halftime (3), and retrograde (4). Depending on the type of transformation, the results vary in their overall quality.
The examples shown in Figure~\ref{results_generate} have been selected to illustrate both high and low quality transformations.
We found that diatonic transposition (TransD) transformation was generally of lower quality than the other transformation types.
In low-quality transformations, notes are frequently missing in generated counterparts, which is a result of learned representations being not fully content-invariant.
Figure~\ref{pca} shows that content-dependence in the first principal component, as this component is not correlated with content-invariant classes.
Note that the GAE is ``conservative'' when it comes to generating analogies, in that the reconstruction errors are caused by omitting existing notes, but very rarely by incorrectly introduced notes.
The reason is that mapping units only take on large values when inputs comply with their transformations \cite{ICML2012Memisevic_105} and as a GAE is symmetric, this also holds for units in the input space.

\section{Conclusion and future work}\label{sec:concl-future-work}
We have evaluated the ability of two unsupervised learning models to learn transformations from pairs of musical n-grams.
We found that the Gated Autoencoder was more effective than a standard RBM, both in an input reconstruction task, and in a discriminative task in which the representations learned by the models were used to classify n-gram pairs in terms of the transformation they exhibit.
Ideally, transformations learned by a GAE are fully content-invariant.
We found that this is not the case in practice when training the GAE using the classical objective function.
In other recent work, we show that the content-invariance of learned representations can be improved by a regularization term that explicitly penalizes content-variance~\cite{lattner17:_improv_conten_invar_gated_autoen_objec_rotat}.

The results reported in this paper show that when given enough n-gram pairs exhibiting transformations of a certain type, the GAE is able to learn these transformations.
Future experiments in which n-gram pairs may exhibit multiple transformation types (for example a diatonic transformation combined with rhythmic changes) should tell to what degree a GAE can disentangle transformations of different types.

The purpose of this evaluation is to assess the suitability of the models for the subsequent task of inferring transformations from a musical corpus.
This poses the further challenge of selecting n-gram pairs from the corpus to train the model on.
Obviously the majority of randomly selected n-gram pairs will be unrelated, and exhibit no meaningful transformations at all.
A possible way to address this issue is to use a bootstrapping approach in which the selection of n-gram pairs for training the model is based on some measure defined by the model itself, for example by greedily selecting the pairs for which the model reconstruction error is lowest.


\section*{Acknowledgments}
This work is supported by the European Research Council (ERC) under the EU's Horizon 2020 Framework Programme (ERC Grant Agreement number 670035, project CON ESPRESSIONE).

\clearpage

\makeatother

\bibliographystyle{named}
\bibliography{bib/bib_mg,bib/bib_cc,bib/bib_sl}

\end{document}